\documentclass[manuscript]{aastex631}
\usepackage{mathrsfs}
\usepackage{amsmath}

\newcommand{\dm}{\mathcal{DM}}

\graphicspath{{./}{figures/}}

\begin{document}

\title{A Catalog of Distance Determinations for the LAMOST DR8 K Giants in the Galactic Halo}

\correspondingauthor{Xiang-Xiang Xue}
\email{xuexx@nao.cas.cn}

\author[0000-0001-7080-0618]{Lan Zhang}
\affiliation{CAS Key Laboratory of Optical Astronomy, National Astronomical Observatories, Chinese Academy of Sciences, Beijing 100101, People’s Republic of China}

\author[0000-0002-0642-5689]{Xiang-Xiang Xue}
\affiliation{CAS Key Laboratory of Optical Astronomy, National Astronomical Observatories, Chinese Academy of Sciences, Beijing 100101, People’s Republic of China}
\affiliation{Institute for Frontiers in Astronomy and Astrophysics, Beijing Normal University, Beijing, 102206, People’s Republic of China}

\author[0000-0003-1972-0086]{Chengqun Yang}
\affiliation{Shanghai Astronomical Observatory, Chinese Academy of Sciences, 80 Nandan Road, Shanghai 200030, People’s Republic of China}

\author{Feilu Wang}
\affiliation{CAS Key Laboratory of Optical Astronomy, National Astronomical Observatories, Chinese Academy of Sciences, Beijing 100101, People’s Republic of China}
\affiliation{School of Astronomy and Space Science, University of Chinese Academy of Sciences, Beijing 100049, People’s Republic of China}

\author[0000-0003-4996-9069]{Hans-Walter Rix}
\affiliation{Max-Planck-Institute for Astronomy K\"{o}nigstuhl 17, D-69117, Heidelberg, Germany}

\author[0000-0002-8980-945X]{Gang Zhao}
\affiliation{CAS Key Laboratory of Optical Astronomy, National Astronomical Observatories, Chinese Academy of Sciences, Beijing 100101, People’s Republic of China}
\affiliation{School of Astronomy and Space Science, University of Chinese Academy of Sciences, Beijing 100049, People’s Republic of China}

\author[0000-0002-1802-6917]{Chao Liu}
\affiliation{School of Astronomy and Space Science, University of Chinese Academy of Sciences, Beijing 100049, People’s Republic of China}
\affiliation{Institute for Frontiers in Astronomy and Astrophysics, Beijing Normal University, Beijing, 102206, People’s Republic of China}
\affiliation{Key Laboratory of Space Astronomy and Technology, National Astronomical Observatories, Chinese Academy of Sciences, Beijing 100101, People’s Republic of China}





\begin{abstract}
We present a catalog of distances for 19, 544 K giants drawn from LAMOST DR8. Most of them are located in the halo of the Milky Way up to $\sim120$~kpc. There are 15\% K giants without SDSS photometry, for which we supplements with Pan-STARRS1 (PS1) photometry calibrated to SDSS photometric system. The possible contamination of the red clumps/horizontal branch are removed according to metallicities and colors before the distance determination. Combining the LAMOST spectroscopic metallicities with the SDSS/PS1 photometry, we estimate the absolute magnitudes in SDSS $r-$band, the distance moduli, and the corresponding uncertainties through an Bayesian approach devised by \citet{xue14} for the SEGUE halo K-giants. The typical distance precision is about 11\%. The stars in the catalog lie in a region of $4-126$ kpc from the Galactic center, of which with 6, 320 stars beyond 20 kpc and 273 stars beyond 50 kpc, forming the largest spectroscopic sample of distant tracers in the Milky Way halo so far.

\end{abstract}



\section{Introduction}
\label{sec:intro} 
K giants, a kind of luminous stars with typical absolute magnitudes of
$-3 < M_r < 1$, are ideal tracers to map the Milky Way halo far beyond
the solar neighborhood. For instance, exploring the formation of the Milky
Way by quantifying the substructures in  the Galactic halo \citep{star09,
xue11,yang19}, estimating the total mass of the Milky Way by kinematics of the tracers
\citep{xue08}, or probing the Milky Way stellar halo profile \citep{xue15,xu18,thom18},
and so on. Good distances and corresponding errors are fundamental to address
such interesting and important questions of our Galaxy,

\citet[][hereafter X14]{xue14} devised a Bayesian approach to estimate the
distances of the Galactic halo K giant, and applied to SEGUE
\citep[the Sloan Extension for Galactic understanding and exploration,][]{yanny09}.
In X14, three priors are considered, which are the stellar number density profile
in the galactic halo, the giant-branch luminosity function, and the different metallicity
distributions of the SEGUE K-giant target subclass. Among them, the prior of the
luminosity function plays the biggest role in the distance estimates. Neglecting it
could cause a systematic bias up to 0.25 mag in distance modulus ($\dm$).
Therefore, the Bayesian approach is optimal to get unbiased distance estimates for stars.

Thanks to the huge number of spectra observed in the Large Sky Area Multi-Object
Fiber Spectroscopic Telescope \citep[LAMOST, also called the Guo Shou Jing Telescope,][]{zhao06, cui12}
and the fact that giants occupy a larger fraction of bright stars, 
generating a large sample of the halo K giants has been made possible. 
In this work, adopting the photometry from SDSS and PS1 (defined in
Section~\ref{sec:data}), we estimate the distances and the uncertainties for a large
sample of K giants observed in the LAMOST survey by following the Bayesian
procedure of X14. Our goal is to present a more complete halo K-giant catalog,
including the intrinsic absolute magnitude $M$, heliocentric distance $D$,
Galactocentric distance $r_{\rm gc}$ and their corresponding errors as well.

This paper is organized as follows. The data used in  this work are introduced in
Section~\ref{sec:data}. The photometry calibration and distance estimation
 are described in Section~\ref{sec:photo_calib}. The results are shown in
Section~\ref{sec:res}, with a summary in Section~\ref{sec:sum}.

\section{Data}
\label{sec:data}
K giants which mainly distributed in the Galactic anti-center direction
\citep{yao12} is covered by the LAMSOT. LAMOST is a
quasi-meridian reflecting Schmidt telescope with an effective aperture of 4
meters and 4000 optical fibers \citep{cui12}, from which numerous low-resolution
($R\sim1800$) spectra covering a wavelength range of $3700 < \lambda < 9000$~{\AA}
of stars with $r < 19$ can be obtained simultaneously at one exposure
\citep{zhao06, zhao12}. 

We aim to derive distance moduli following the procedure of X14 for K-giant
halo stars selected from the LAMOST DR8. With this method, the absolute magnitude
of each star is estimated from the empirically calibrated color-magnitude fiducials
with metallicities in the range of $-2.38 < {\rm [Fe/H] < +0.39}$. The main observables
adopted are [Fe/H], $\log g$ and $T_{\rm eff}$ derived from LAMOST stellar spectra,
the $g$ and $r$ band magnitudes from the Sloan Digital Sky Survey \citep[SDSS,][]{york00}
or the Pan-STARRS1 survey \citep[PS1,][]{cham16}. 
In addition, to eliminate possible
red clump stars, 2MASS \citep{skru06} $J$ and $K_{\rm s}$ band magnitudes
are also used. Details of the observables are described below.

Stellar atmospheric parameters are derived by the official
LAMOST Stellar Parameter pipeline \citep[LASP;][]{wu11, luo15}, in which the stellar
parameters are determined iteratively by minimizing the $\chi^2$
between the observed spectrum and the model spectrum from ELODIE
stellar library \citep{prug07}. From all the survey spectra, the K giants used in the study
were identified by the selection criteria presented in \citet[][Figure~3]{liu14}, that is,
$4000 < T_{\rm eff} < 5600$~K, and $\log g < 3.5$ for stars whose $T_{\rm eff} < 4600$~K,
while $\log g < 4.0$ for stars with $4600 \leq T_{\rm eff} < 5600$~K. It returns 
$\sim 0.98$~million K giants in LAMOST DR8, including $\sim 16\%$ red clump (RC) giant
by using the criterion of \citet{huang15}, that is,
\begin{equation}
\label{eq:rc1}
	1.8 \leq \log g \leq 0.0009\, {\rm dex\, K^{-1}} \{T_{\rm eff} - T^{\rm Ref}_{\rm eff} ({\rm [Fe/H]})\} + 2.5,
\end{equation}
where
\begin{equation}
\label{eq:rc2}
	T^{\rm Ref}_{\rm eff} ({\rm [Fe/H]}) = -876.8\, {\rm K \, dex^{-1}\, [Fe/H]} + 4431\, {\rm K},
\end{equation}
and
\begin{equation}
\label{eq:rc3}
	1.21 [(J - K_{\rm s})_0 - 0.085]^9 + 0.0011 < Z_{\rm metal} < 2.58 [(J - K_{\rm s})_0 - 0.400]^3 + 0.0034,
\end{equation}
where $Z_{\rm metal}$ is converted from [Fe/H] using Eq.~10 of \citet{bert94}, and $(J - K_{\rm s})_0$
are photometries taken from 2MASS.
            
Photometries are firstly adopted from SDSS which use the 2.5~m Sloan Foundation
Telescope \citep{gunn06} at Apache Point Observatory (APO). The 16th
public data release \citep[DR16;][]{sdss16} was from the SDSS-IV
\citep{blan17}, which contains all prior
SDSS $ugriz$ imaging data. 
\citep{fuku96, gunn98, york00, stou02, pier03, eise11, blan17}.
For K giants without SDSS photometry, we took the broadband
data $grizy_{\rm P1}$ from the second data release
\citep[DR2;][]{pan16, pan20a, pan20b, pan20c, pan20d}
of PS1 instead, while the Panoramic Survey Telescope and Rapid Response System (Pan-STARRS)
is an innovative wide-field astronomical imaging and data processing
facility developed at the University of Hawaii’s Institute for
Astronomy \citep{kais02, kais10}. 
The DR2 of PS1 data used in the present work can be found in
MAST: \dataset[10.17909/s0zg-jx37]{https://doi.org/10.17909/s0zg-jx37}.
Finally, there are 45\% and  83\% stars
with SDSS and PS1 photometric data, respectively.
Moreover, to select halo K giants with good data quality, only stars whose
$E(B-V)$ estimated from \citet{schl98} and less than 0.25 mag. are adopted.
Hence, there are 57\% stars left.
Then the extinctions in different bandpass were calculated by \citet{f99}
reddening law with measurements of \citet{schl11}.

\section{Photometry Calibration and Distance Estimates}
\label{sec:photo_calib}
\begin{figure} [ht!]
\centering
\includegraphics[scale=0.50]{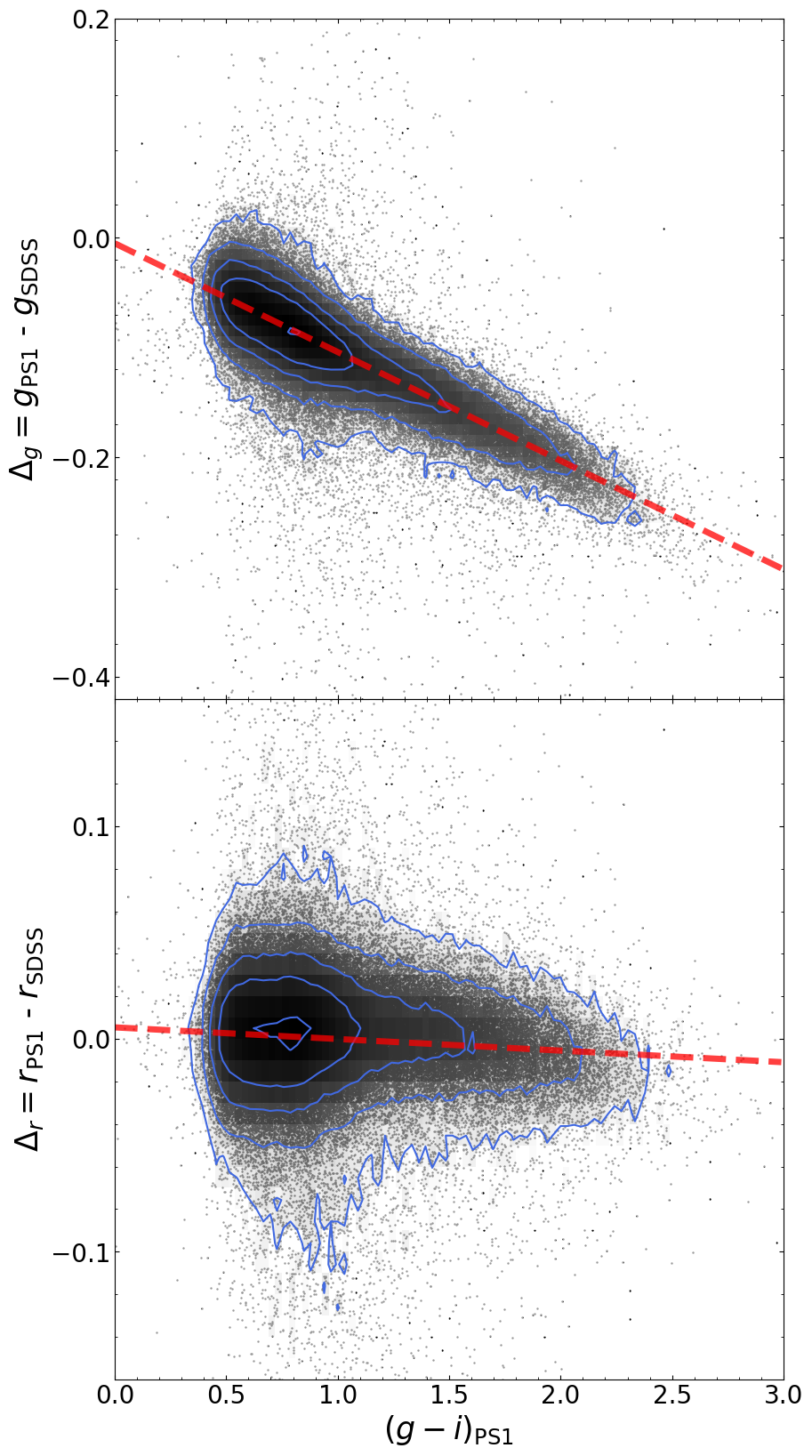}
\caption{ Apparent magnitude difference between PS1 and SDSS as functions of PS1
	    color $(g - i)_{\rm PS1}$ for the mags of $g$ and $r$.  Density distributions
	    in $\Delta$ -- color spaces are shown. Red dashed lines represent the best
	    fit to the data.
	    \label{fig:photo_cali}}
\end{figure}

For stars without SDSS photometry, the PS1 data is taken as a complement. From 
common stars both observed by SDSS and PS1, trends in the difference between
the PS1 and SDSS photometry $\Delta_m =  m_{\rm PS1} - m_{\rm SDSS}$ with
PS1 colors $(g - i)_{\rm PS1}$ can be found (see Figure~\ref{fig:photo_cali}).
Thus, we need to remove the trends by calibrating the PS1 photometry to match
the SDSS one before distance estimates.

Here we perform the calibration by fitting an empirical curve to $\Delta_m$,
where $m$ represents apparent magnitudes $g$ and $r$, as a function of PS1
color $(g - i)$, and thus adjusting the PS1 photometry. To model
$\Delta_m - (g - i)$ precisely, we chose stars both observed by SDSS and PS1 in
the apparent magnitude ranges of $13. < g_{\rm SDSS} < 22.2$, $13. < r_{\rm SDSS} < 22.2$,
$13. < i_{\rm SDSS} < 21.3$, $13. < g_{\rm PS1} < 22$, $13. < r_{\rm PS1} < 21.8$, and
$13. < i_{\rm PS1} < 21.5$, with small observation uncertainties, that is,
$\sigma_m < 0.05$ ($m = g, r, i$). It results 168, 887 common stars used for the
present calibration.

\begin{figure}. 
\gridline{\fig{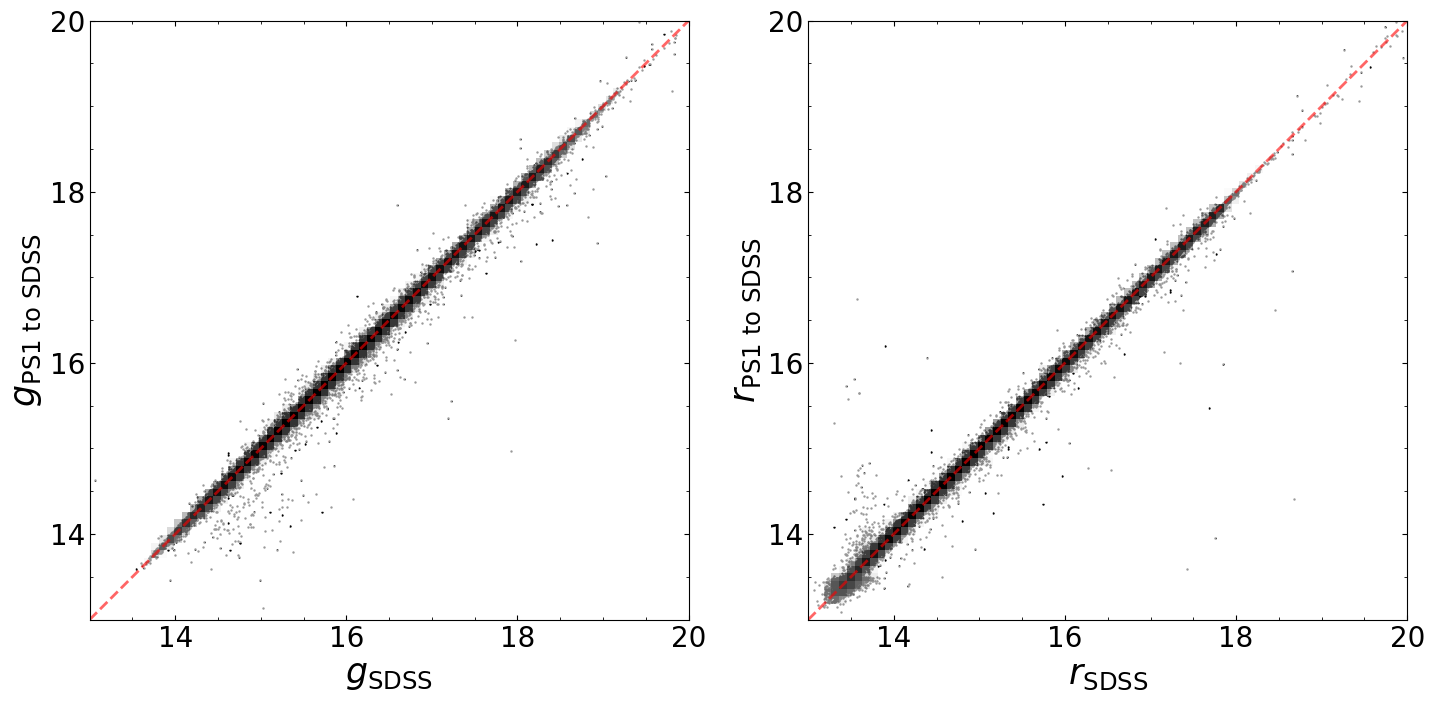}{\textwidth}{(a)}}
\gridline{\fig{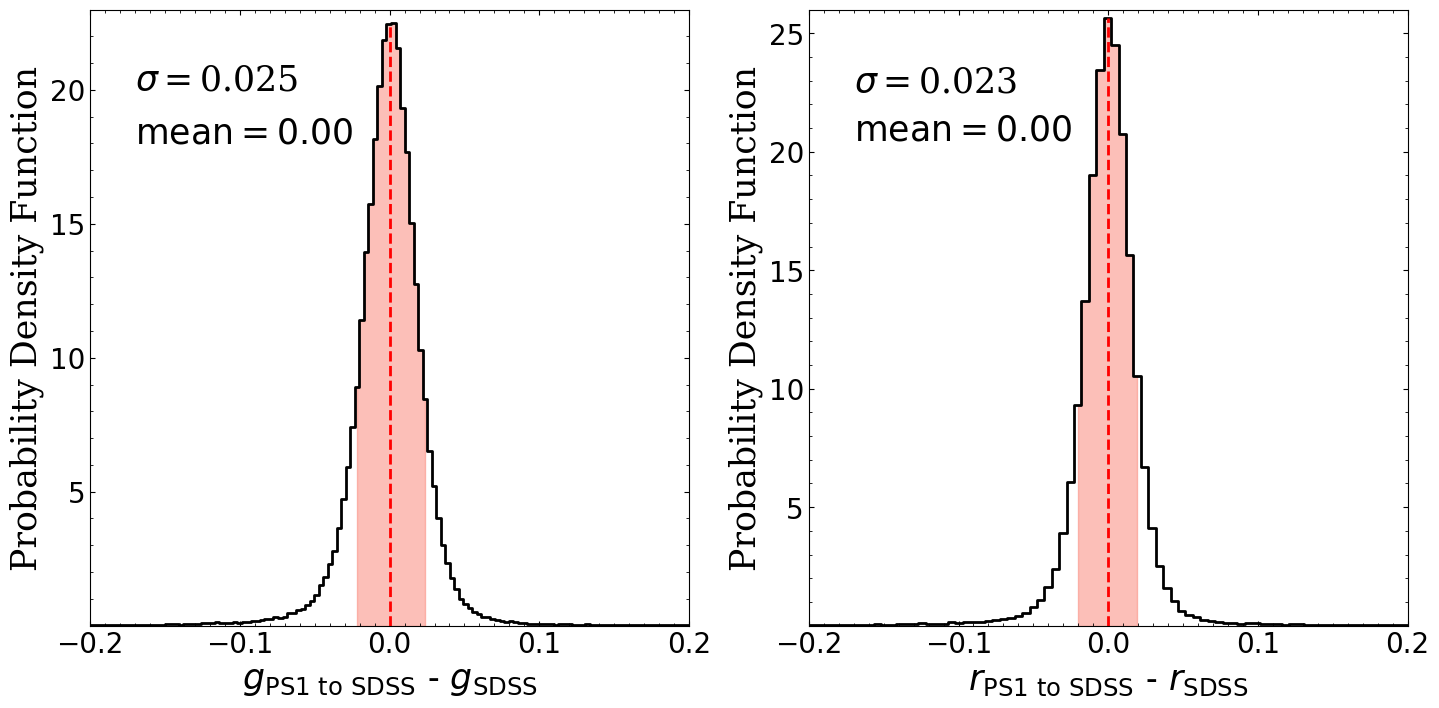}{\textwidth}{(b)}}
\caption{Calibration results. The upper panels show adjusted PS1 photometry. Red
	   dashed lines are diagonals. The lower panels present PDFs of $\Delta_m$
	   after calibration. The mean values and 68\% intervals are shown in red
	   lines and shades, respectively.
	   \label{fig:ps2sdss}}
\end{figure}

We find the best fit slopes of lines in $\Delta_m$ vs. $(g - i)_{\rm PS1}$ space
with least absolute deviation line fitting, which minimizes the absolute value of
the residuals,
\begin{equation}
\label{eq:lad}
	\sum_j | \Delta_{m, j} - F_m((g - i)_{{\rm PS1}, j})|,
\end{equation}
to exclude outliers. The standard errors of the fitted slopes are estimated via
bootstrapping. Figure~\ref{fig:photo_cali} shows the best fits in 
$\Delta_m - (g - i)_{\rm PS1}$ space, the fittings are:
\begin{equation}
\label{eq:cali_g}
\begin{aligned}
	\Delta_g  = &\,\, (-0.0050 \pm 1.0 \times 10^{-4}) \, + \, \\
	                    & \,\,(-0.0099 \pm 7.1 \times 10^{-5}) \times (g - i)_{\rm PS1},
\end{aligned}
\end{equation}
and 
\begin{equation}
\label{eq:cali_r}
\begin{aligned}
	\Delta_r = &\,\, (0.0054 \pm 2.1 \times 10^{-4}) \, + \, \\
	                  &\,\, (-0.0055 \pm 1.5 \times 10^{-4}) \times (g - i)_{\rm PS1}.
\end{aligned}
\end{equation}
The calibrated results are shown in Figure~\ref{fig:ps2sdss}. The upper panels
present that the calibrated $g_{\rm PS1}$ and $r_{\rm PS1}$ show very good
consistency with the SDSS photometry. In the lower panels,  it can be seen that
the  mean values of adjusted $\Delta_m$ are 0.0 with standard errors of 0.025
and 0.023 for $g$ and $r$ mags, respectively, which proves the validity of
the calibration in this work.

\begin{figure}[ht!]
\centering
\includegraphics[scale=0.50]{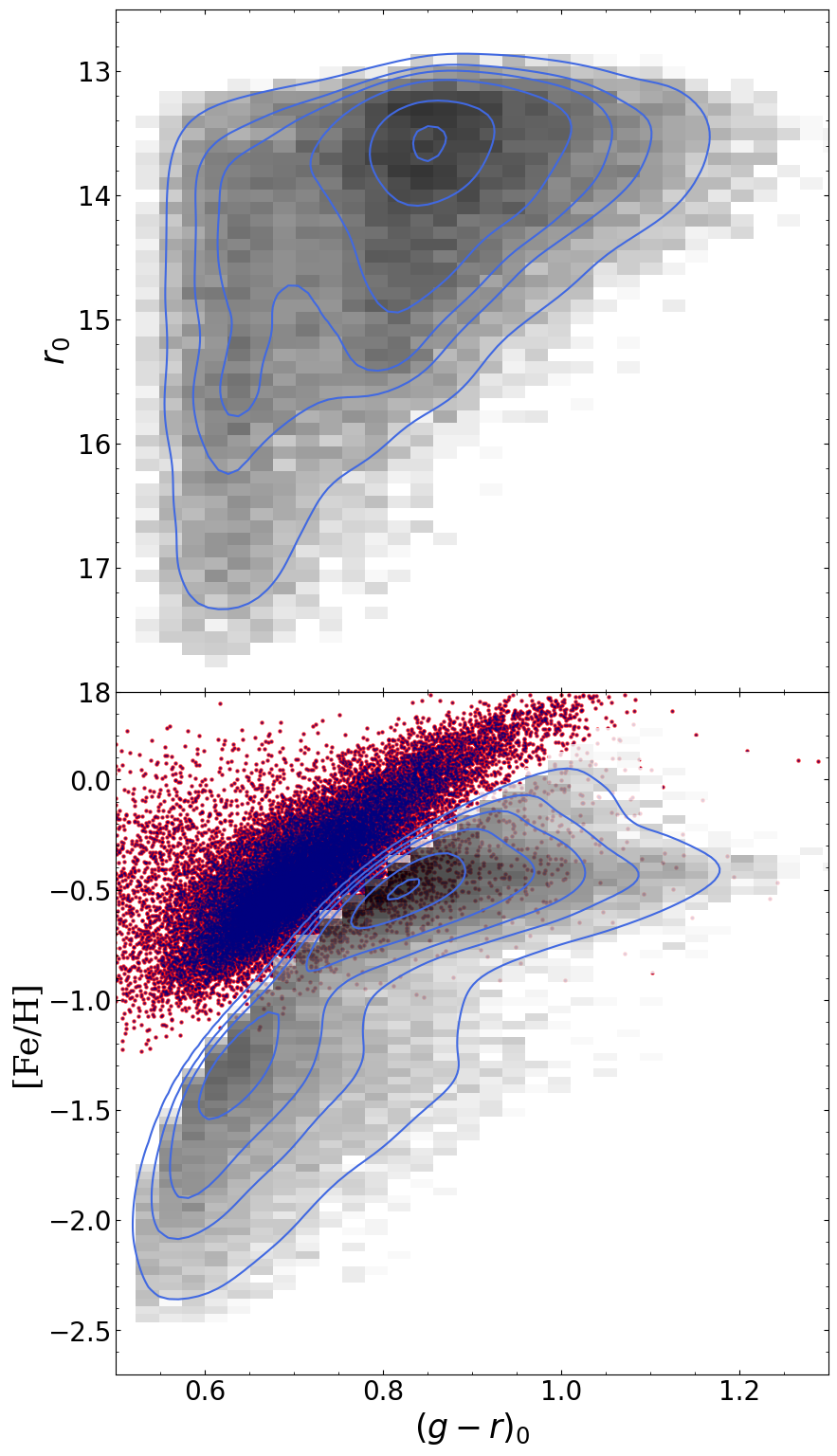}
\caption{Number density distributions of K-giant sample in the
	   color-magnitude and color-metallicity plane. In the bottom panel, red
	   and dark blue points represent possible RC and HB stars, respectively.
	   \label{fig:sample}}
\end{figure}

With the calibrated results, we converted PS1 photometries to SDSS ones
for stars without SDSS observation. Finally, we obtains 375, 585 stars with $g_0$, $r_0$, $i_0$, and 
$E(B-V) < 0.25$. The present distance measurement is for halo stars only. 
Therefore, we selected halo K giants preliminarily before the distance measurements.
Considering the colors and metallicities of used giant-branch fiducials of clusters, only K giants whose
${\rm [Fe/H] < +0.39}$ and $0.5 \leq (g - r)_0 \leq 1.4 $ are taken.
We further excluded stars labeled as RC and stars below the level of the horizontal branch (BH)
by using a quadratic polynomial of $(g - r)_0^{\rm HB} = {\rm 0.087 [Fe/H]^{2} +0.39 [Fe/H] +0.96}$,
which is fitted by X14, 
since the red HB as well as the RC in a cluster have the same color as
K giants, but quite different absolute magnitudes, which in the end leaves 42, 713 stars. 
Figure~\ref{fig:sample} shows
the number density distribution of K giants in $r_0$ vs. $(g - r)_0$ and
[Fe/H] vs. $(g - r)_0$ spaces. It can be notice that partial RC stars overlap with
K giants, since the Galactic halo contains metal-rich substructures\citep{yang19}.

We then implemented a Bayesian approach to derive the posterior PDF of $\dm$
for each K giant, and hence to provide both a distance estimate and its
uncertainty. According to X14, the relative probability of
different $\dm$ is defined as
\begin{equation}
 \label{eq:ppdf}
 \begin{aligned}
 P(\dm | \{m, c, {\rm [Fe/H]}\}) = & \frac{P(\{m, c, {\rm [Fe/H]}\} | \dm)}{P(\{m, c, {\rm [Fe/H]}\})} \\
                                                       & \times p_{\rm prior}(\dm),
\end{aligned}
\end{equation}
where $P(\{m, c, {\rm [Fe/H]}\})$ is a non-zero constant of normalization,
$p_{\rm prior}(\dm)$ is the prior probability for the $\dm$. It reflects any
information on the stellar number density which follows a power law
$\nu(r) = r^{\alpha}$ in the Galactic halo, where $\alpha \in (-4, -2)$
\citep{bell08}. With mock data, X14 has proved that the variances in the final
estimated $\dm$ caused by $p_{\rm prior}(\dm)$ can be neglected if $\alpha$
values are in $(-4, -2)$. $P(\{m, c, {\rm [Fe/H]}\} | \dm)$ is the likelihood
of $\dm$, e.g., $\mathscr{L} (\dm)$, in which {\it prior} information of the
giant-branch luminosity function $p_{\rm prior} (M)$ and the metallicity
distribution of halo giants $p_{\rm prior} ({\rm [Fe/H]})$ are involved, that is,
\begin{equation}
 \label{eq:likeli}
 \begin{aligned}
 \mathscr{L} (\dm) = & \int \int p(\{m, c, {\rm [Fe/H]} | \dm, M, {\rm [Fe/H]}\}) \\
 		          & \times p_{\rm prior}(M) p_{\rm prior}({\rm [Fe/H]}) dM d{\rm [Fe/H]}.
 \end{aligned}
\end{equation}
$p_{\rm prior}(M) \propto 10^{0.32M}$ is derived from the observed luminosity
functions of the giant branch of global clusters M~5
\citep[${\rm [Fe/H]} = -1.4$,][]{sand96} and M~30
\citep[${\rm [Fe/H]} = -2.13$,][]{sand99}, and the Basti theoretical ones with
${\rm [Fe/H]} = -2.4$ and ${\rm [Fe/H]} = 0.0$ \citep{piet04}. 
$p_{\rm prior} ({\rm [Fe/H]})$ is from metallicity distribution of the K-giants.

To calculate $\mathscr{L} (\dm)$,  the colors at a given $M$ and [Fe/H],
$c(M, {\rm [Fe/H]})$,  need to be predicted first. The value is interpolated
from a set of empirical giant-branch fiducials of $(g - r)_0$ - $M_r$, which
adopted from the globular clusters M~92, M~13, and M~71, the open cluster
NGC~6791, and the Basti $\alpha-$enhanced isochrones \citep{piet06}.
$\alpha$ abundance also plays an important role in $c(M, {\rm [Fe/H]})$ determination,
e.g., at the same [Fe/H] values, the difference of $r$-band absolute magnitude
can be as large as 0.5~mag at the tip of the giant branch between an
$\alpha$-enhanced giant and a solar-scaled $\alpha$-abundance one (X14).
However, in the Galactic halo, a few K giants are metal-rich with poor
[$\alpha$/Fe], such as K giants which are belong to Sagittarius Stream
\citep{yang19}. In this situation, for the K giants with
$0. < {\rm [Fe/H]} < +0.39$,  the $\alpha$ abundances are assumed to be gradually
weakening between solar and the NGC~6791 value, while for normal $\alpha$-enhanced
metal-poor halo stars, the cluster fiducial and one isochrone are used. 

For more details of the distance calculation procedure, we refer the readers to X14,
and reference therein. Finally, the best estimates of the distance moduli and their
errors are given by the median and central 68\% interval of $\mathscr{L} (\dm)$.

\section{Results}
\label{sec:res}
\begin{figure}[ht!]
\centering
\includegraphics[width=\textwidth]{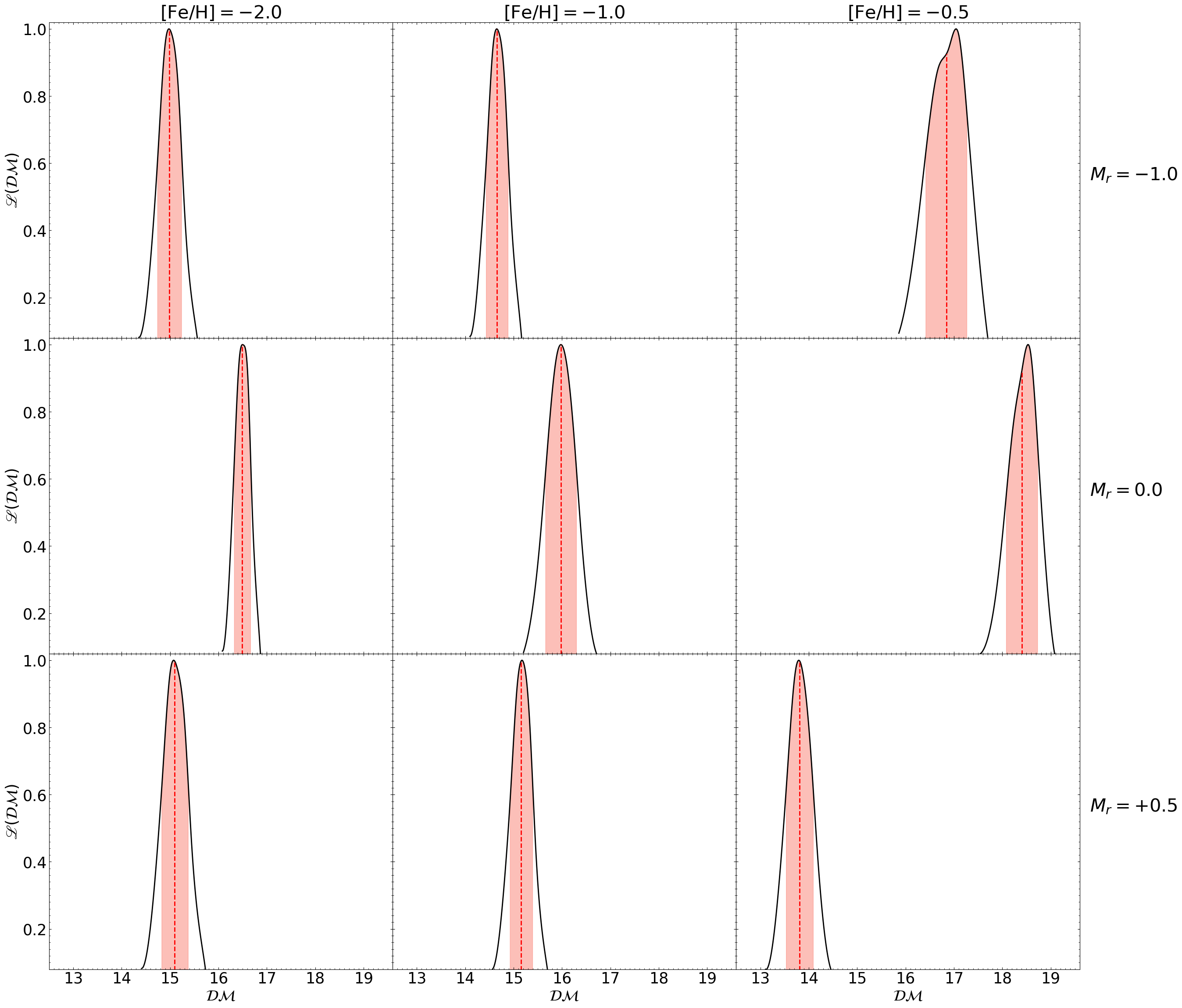}
\caption{Typical $\mathscr{L} (\dm)$ for stars with different [Fe/H] and $M_r$.
	   In each plot, the black thick line represents the most likely $\dm$. 
	   The red dashed line and the shade indicate the median value and 68\%
	   interval, respectively.
	    \label{fig:l_dm}}
\end{figure}

Figure~\ref{fig:l_dm} shows examples of the $\dm$ estimates and corresponding
uncertainties from $P(\dm | \{m, c, {\rm [Fe/H]}\})$. Meanwhile, the absolute magnitudes,
$M_r ( = r_0 - \dm_{\rm median})$, distances ($D$) from the Sun,
Galactocentric distances ($r_{\rm gc}$), positions ($[x, y, z]$) in the Galacticcentric
Cartesian coordinate system by adopting $R_{\odot}= 8.34$~kpc \citep{reid14}
are obtained. To generate reliable catalog for the Galactic halo K giants, we first
exclude disk stars and the K giants with bad distance estimates through the criteria of
$|z| > 5.$~kpc, ${\rm [Fe/H]} < -1.0$ if $2.~{\rm kpc} < |z| < 5.$~kpc, and
$\sigma_{\rm D} / {\rm D} < 0.3$. Then we calculate the kinematic parameters by combining
proper motions from Gaia Early Data Release 3 \citep[EDR3;][]{gaia16, gaia21},
and hence the total energy for the rest stars to make sure the given K giants are bound to
the Milky Way, which weeds out 665 stars. Finally, It results a catalog of
19, 544 halo K giants. The main entries in the catalog are
the best estimates of the distance moduli and their uncertainties ($\dm_{\rm median}$, $\sigma_{\dm}$),
the heliocentric distances and their errors (${\rm D}$, $\sigma_{\rm  D}$),
the Galactocentric distances and their errors ($r_{\rm gc}$, $\sigma_{r_{\rm gc}}$),
the absolute magnitudes in $r-$band and the corresponding errors ($M_r$, $\sigma_{M_r}$),
and the $\dm$ at  (5\%, 16\%, 50\%, 84\%, 95\%) confidence of $\mathscr{L} (\dm)$.
These values and other parameters are all included in Table~\ref{tab:dis_res}.
The complete version of this table is available in the online version.

\begin{figure}[ht!]
\gridline{\fig{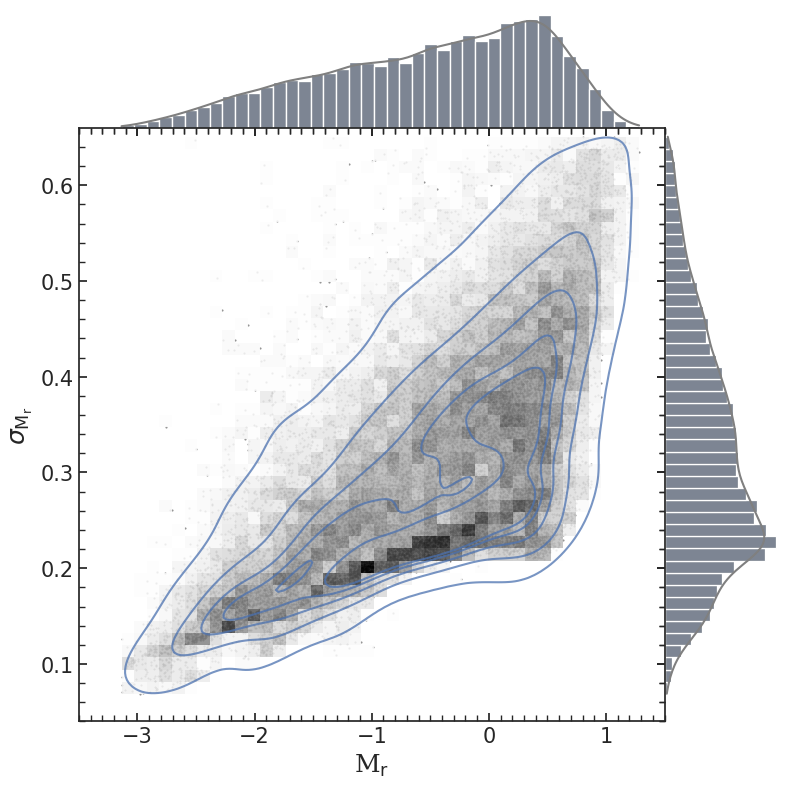}{0.42\textwidth}{}}
\vspace{-0.8cm}
\gridline{\fig{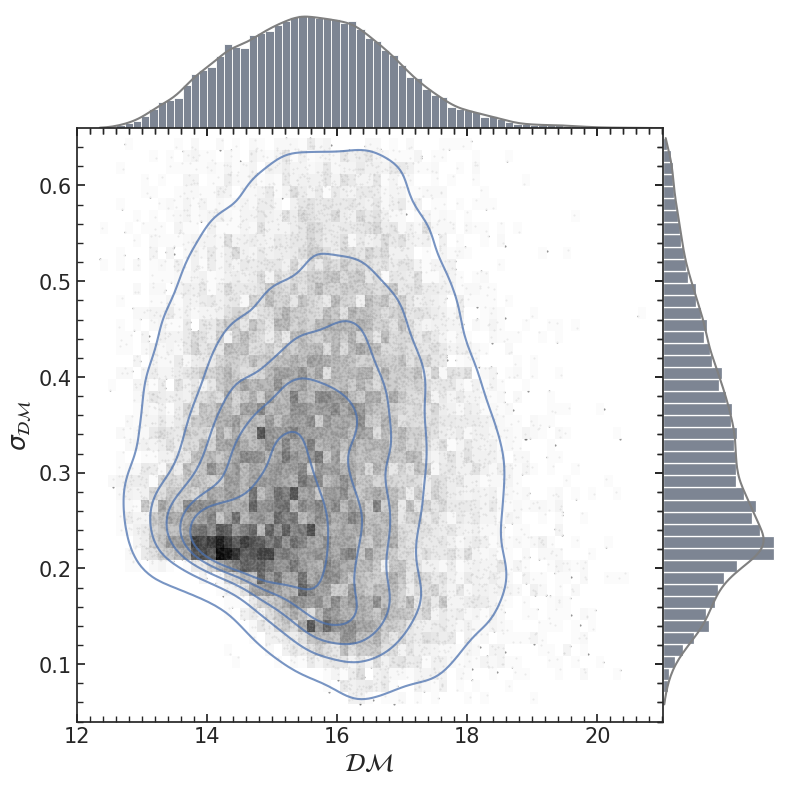}{0.42\textwidth}{}}
\vspace{-0.8cm}
\gridline{\fig{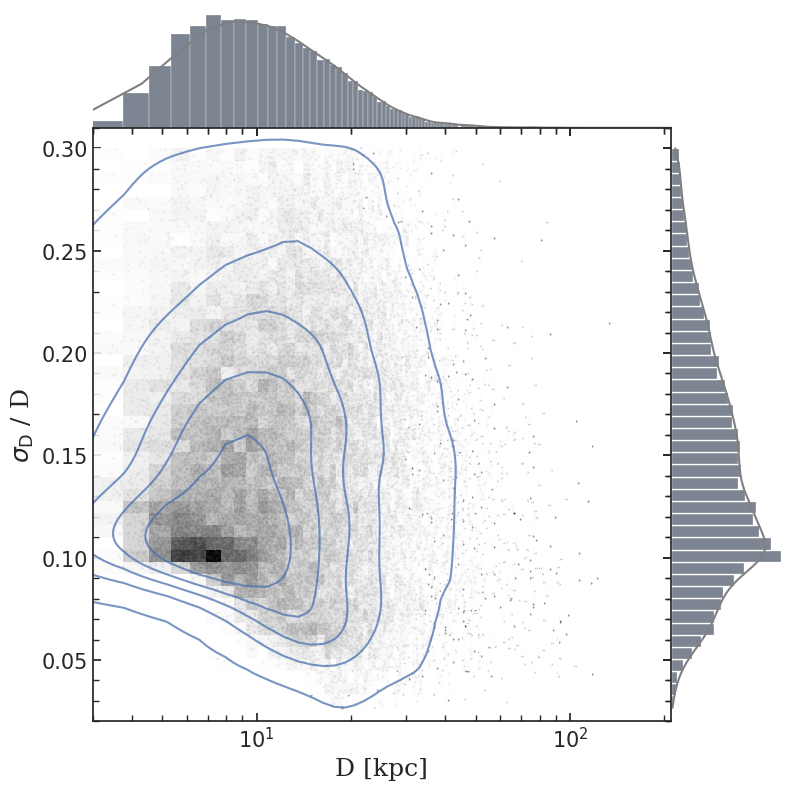}{0.42\textwidth}{}}
\caption{Distance results for 19, 544 halo K giants. The density distribution
	     maps of the absolute magnitudes in $r-$band, $\dm$, and distance vs.
	     their corresponding uncertainties are presented from upper to bottom
	     panels, respectively.
	     \label{fig:dis_err}}
\end{figure}

\begin{figure}[ht!]
\centering
\includegraphics[width=\textwidth]{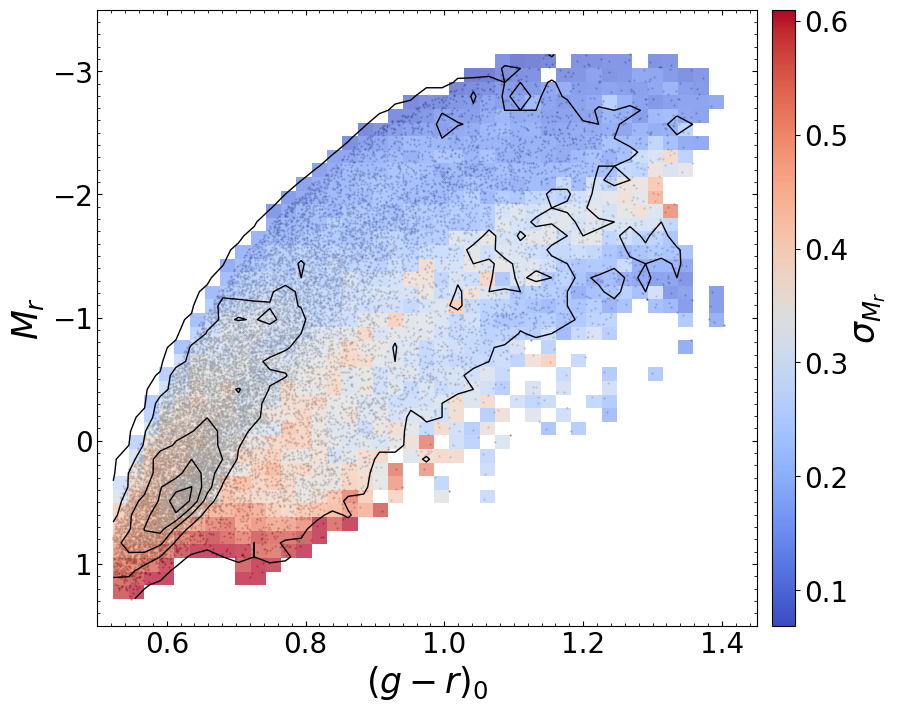}
\caption{The errors of the absolute magnitude of $r-$band,  $\sigma_{M_r}$ as a 
	   function of $M_r$ and $(g - r)_0$. Black contours are the density distribution
	   of $M_r - (g - r)_0$ plane.
	   \label{fig:cmd}}
\end{figure}

\begin{figure}[ht!]
\centering
\includegraphics[width=\textwidth]{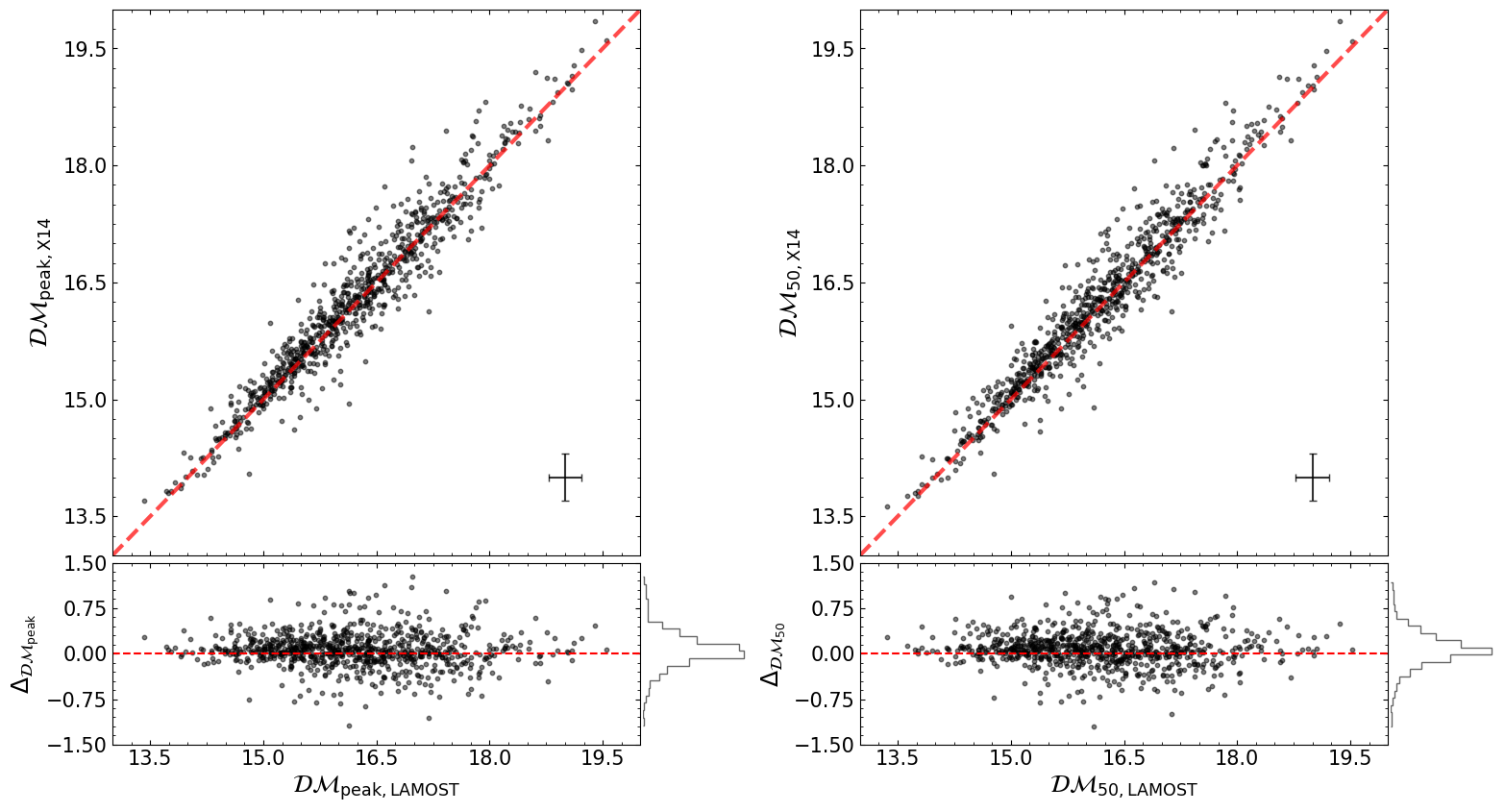}
\caption{Distance comparison for 795 common stars between the present catalogue and the one
	     in X14. The peak values, and $\dm$ at 50\% of $\mathscr{L} (\dm)$ are shown in the left
	     and right panels, respectively. Diagonals and values of $\Delta_{\dm} = 0.$ are
	     shown in red dashed lines. Note that
	     $\Delta_{\dm}=\dm_{\rm X14} - \dm_{\rm LAMOST}$ for both of the lower panels.
	   \label{fig:common}}
\end{figure}

The overall properties of the ensemble of distance estimates are shown in
Figure~\ref{fig:dis_err}. It can be seen that the average precisions in both of
$M_r$ and $\dm$ are 0.32 mags, and the faint giants have less precise in $M_r$
estimates. The mean relative error in distance, $\sigma_{\rm D} / {\rm D}$, is 0.145,
and the present K giants can reach as far as 120~kpc with very good precisions
(e.g., $\sigma_{\rm D} / {\rm D} < 0.15$). The distance results of fractional nearby
stars are less precise  because they tend to be on the steep part of the giant branch
in the CMD, especially for stars in low metallicity range. Figure~\ref{fig:cmd} shows the
number density and $\sigma_{M_r}$ distribution of halo K giants in the CMD.
More stars gather in the lower part of the giant branch, which is consistent with the
prediction of the luminosity function. The precision of the distance is increasing
along with the red-giant branch, that is, stars in the upper part of the RGB have more
precise distances than those in the lower part because the fiducials are much steeper
near the sub-giant branch. This characters is consistent with the distance estimates
of SEGUE K giants in X14. Moreover, we compare the present estimates with ones
in X14 for 795 common stars in Figure~\ref{fig:common}. In this figure, both peak values
($\dm_{\rm peak}$) and $\dm$ at 50\% of $\mathscr{L} (\dm)$ are shown. The standard
deviations are 0.29 and 0.28 for $\Delta_{\dm_{\rm peak}}$ and $\Delta_{\dm_{50}}$, respectively.
It can be seen that the two estimates show good consistency and the $\Delta_{\dm}$ values are
independent of the estimated $\dm$, which proves the applicability of
this approach to the LAMOST data. The scatter is mainly caused by different input
[Fe/H] values used in the distance estimates.

\begin{figure}[ht!]
\centering
\includegraphics[width=\textwidth]{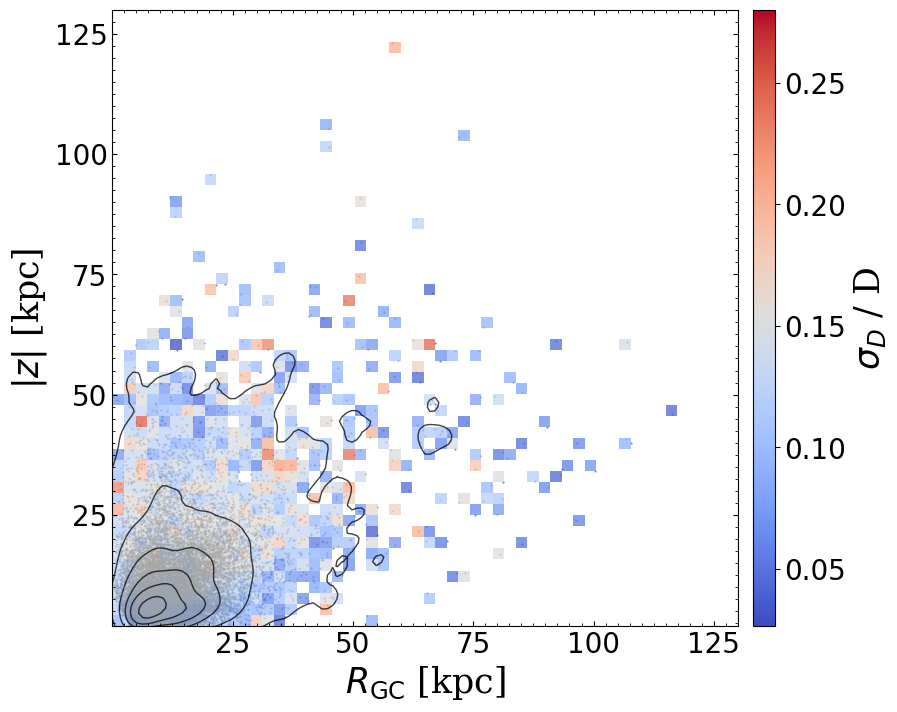}
\caption{The number and $\sigma_{\rm D}$ / D distributions in $R_{\rm GC} - |z|$
	    plane for the halo K giants. The stars can trace as far as
	    $R_{\rm GC} = 125$~kpc with good precisions.
	   \label{fig:Rz}}
\end{figure}

Figure~\ref{fig:Rz} presents the number density and $\sigma_{\rm D} / {\rm D}$ in
$R_{\rm GC} - |z|$ plane. Our sample of K giants lies in the region of $4 - 126$ kpc from
the Galactic center. In the sample, 6, 320 stars are in the region of 
$r_{\rm gc} > 20$~kpc, including 221 stars in $52 < r_{\rm gc} < 80$~kpc and
68 stars  whose $r_{\rm gc}$ are larger than 80~kpc. The sample size is as double as
SEGUE K giants in X14. The large sample can help us to trace the mass density for both
of the Galactic inner and outer halos in the following works.

\begin{figure}[ht!]
\centering
\includegraphics[width=0.75\textwidth]{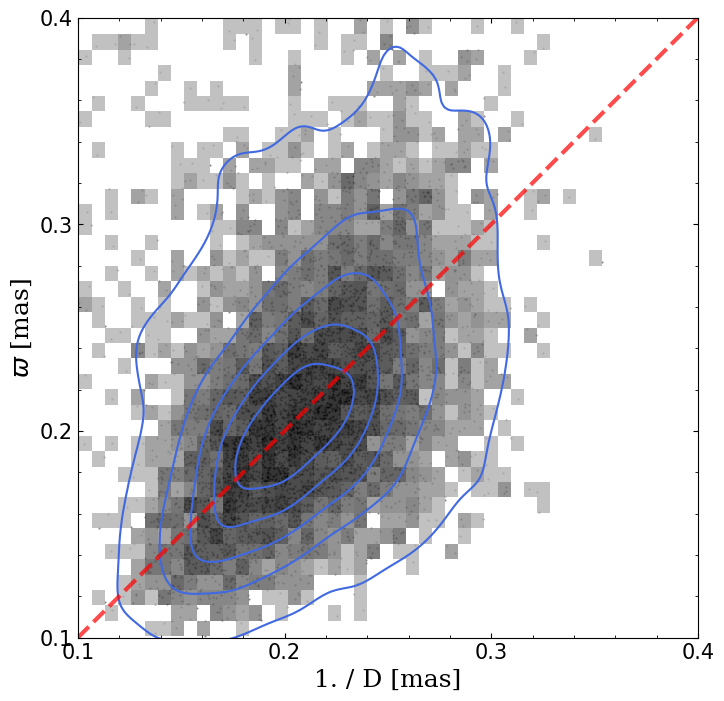}
\caption{Our distances of good estimates ($\sigma_{\rm D} / {\rm D} < 0.15$)
	      compare with parallaxes whose relative errors $\sigma_{\varpi} / \varpi < 0.1$
	      from Gaia EDR3. Red dashed lines are diagonals.
	   \label{fig:dis_com}}
\end{figure}

To check the validation of our distance estimates in this work, we also compare our
results with the parallaxes adopted from Gaia EDR3. Only stars in
$\sigma_{\varpi} / \varpi < 0.1$ and $\sigma_{\rm D} / {\rm D} < 0.15$ are taken,
which results 5, 497 K giants. The comparison result is shown in Figure~\ref{fig:dis_com},
in which our distance estimates show very good consistency with parallaxes from Gaia
EDR3 catalog for stars with precise parallax measurements.

\section{Summary}
\label{sec:sum}
We have selected the K giants from LAMSOT DR8, excluding red clumps and stars
below the horizontal branch carefully. After calibrating the PS1 photometry
$gi_{\rm PS1}$ to the $gr_{\rm SDSS}$ of SDSS DR16, we combined the $gr$ with the
[Fe/H] from LAMOST LASP, to estimate intrinsic absolute magnitudes in $r-$band,
$\dm$, the distances, and the corresponding uncertainties through an Bayesian approach
for the K giants. The priors adopted in this work are the giant-branch luminosity
function derived from globular clusters, and the metallicity distributions from the present K-giant
sample. The predicted colors $c(M_r, {\rm [Fe/H]})$,  are obtained from empirical
color-magnitude fiducials from old stellar clusters, the best estimates of the distance moduli
and their errors can be estimated using the median value and central 68\% interval of
$\mathscr{L} (\dm)$. The mean relative distance error is 0.145, and the distance errors of
stars which lie on the lower part of the giant branch in the CMD are relative larger.
With the estimated distances and their errors, we selected halo K giants
in the ranges of $|z| > 5.$~kpc, ${\rm [Fe/H]} < -1.0$ if $2.~{\rm kpc} < |z| < 5.$~kpc, and
$\sigma_{\rm D} / {\rm D} < 0.3$. After excluding stars are not bound by the Galactic potential,
it results 19, 544 halo K giants, which is three times larger than the number of X14 from SEGUE
K giants. In the catalog, 6, 320 stars are in the region of  $r_{\rm gc} > 20$~kpc, including 221
stars in $50 < r_{\rm gc} < 80$~kpc and 52 stars  whose $r_{\rm gc}$ are larger than 80~kpc.
Furthermore, we compared the distances  with parallaxes from Gaia EDR3 catalog for stars
with precise parallax measurements. The result presents a very good consistency, which
proves the validation of the estimates in this work.

We finally present an online catalog containing the $\dm$, and LASP atmospheric
parameters for the 19, 544 halo K giants. For each object in the catalog, we also list
the basic observables such as (R.A., Dec.), extinction corrected apparent magnitudes
and de-reddened colors, and heliocentric radial velocities from LAMOST. The Bayesian
estimates of the $\dm$, heliocentric distance, Galactocentric distances, the absolute
magnitudes and their uncertainties, along with the $\dm$ at 
[5\%, 16\%, 50\%, 84\%, 95\%] confidence of $\mathscr{L} (\dm)$.

\begin{acknowledgments}
This work is supported by National Key Research and Development Program of China No. 2019YFA0405500,
National Natural Science Foundation of China (NSFC) under grants No. 11988101, 11890694, 11873052, and 
China Manned Space Project with No. CMS-CSST-2021-B03. L.Z. and X-X.X. acknowledge the support
from CAS Project for Young Scientists in Basic Research Grant No. YSBR-062. F.-L.W. acknowledge
the support from NSFC under grants No. 12073043.

Guoshoujing Telescope (the Large Sky Area Multi-Object Fiber Spectroscopic
Telescope LAMOST) is a National Major Scientific Project built by the Chinese
Academy of Sciences. Funding for the project has been provided by the National
Development and Reform Commission. LAMOST is operated and managed by the
National Astronomical Observatories, Chinese Academy of Sciences. 

Funding for the Sloan Digital Sky Survey (SDSS) has been provided by the
Alfred P. Sloan Foundation, the Participating Institutions, the National
Aeronautics and Space Administration, the National Science Foundation, the
U.S. Department of Energy, the Japanese Monbukagakusho, and the Max Planck
Society. The SDSS Web site is http://www.sdss.org/.

The SDSS is managed by the Astrophysical Research Consortium (ARC) for the
Participating Institutions. The Participating Institutions are The University
of Chicago, Fermilab, the Institute for Advanced Study, the Japan Participation
Group, The Johns Hopkins University, Los Alamos National Laboratory, the
Max-Planck-Institute for Astronomy (MPIA), the Max-Planck-Institute for 
Astrophysics (MPA), New Mexico State University, University of Pittsburgh,
Princeton University, the United States Naval Observatory, and the University
of Washington.

The Pan-STARRS1 Surveys (PS1) and the PS1 public science archive have been made
possible through contributions by the Institute for Astronomy, the University of
Hawaii, the Pan-STARRS Project Office, the Max-Planck Society and its participating
institutes, the Max Planck Institute for Astronomy, Heidelberg and the Max Planck
Institute for Extraterrestrial Physics, Garching, The Johns Hopkins University,
Durham University, the University of Edinburgh, the Queen's University Belfast,
the Harvard-Smithsonian Center for Astrophysics, the Las Cumbres Observatory
Global Telescope Network Incorporated, the National Central University of Taiwan,
the Space Telescope Science Institute, the National Aeronautics and Space
Administration under Grant No. NNX08AR22G issued through the Planetary Science
Division of the NASA Science Mission Directorate, the National Science Foundation
Grant No. AST-1238877, the University of Maryland, Eotvos Lorand University (ELTE),
the Los Alamos National Laboratory, and the Gordon and Betty Moore Foundation.

This work has made use of data from the European Space Agency (ESA) mission
{\it Gaia} (\url{https://www.cosmos.esa.int/gaia}), processed by the {\it Gaia}
Data Processing and Analysis Consortium (DPAC,
\url{https://www.cosmos.esa.int/web/gaia/dpac/consortium}). Funding for the DPAC
has been provided by national institutions, in particular the institutions
participating in the {\it Gaia} Multilateral Agreement

\end{acknowledgments}

\bibliography{ref}{}
\bibliographystyle{aasjournal}

\clearpage

\clearpage

\begin{splitdeluxetable}{ccrrrrrcrrrrBrrrrrrrrrrrrrr}
\tablewidth{\textwidth} 
\tablenum{1}
\tablecaption{Catalog of 19, 544 K Giants Selected from LAMOST DR8 \label{tab:dis_res}}
\tablehead{
\colhead{R.A. (J2000)} & \colhead{Dec. (J2000)} & \colhead{$r_0$} & \colhead{$\sigma_{r_0}$} &
\colhead{$(g - r)_0$} & \colhead{$\sigma_{(g - r)_0}$} & \colhead{RV} & \colhead{$\sigma_{\rm RV}$} &
\colhead{$T_{\rm eff}$} &\colhead{$\sigma_{T_{\rm eff}}$} & \colhead{$\log g$} & \colhead{$\sigma_{\log g}$} &
\colhead{[Fe/H]} & \colhead{$\sigma_{\rm [Fe/H]}$} & \colhead{$\dm_{5\%}$} & \colhead{$\dm_{16\%}$} &
\colhead{$\dm_{50\%}$} & \colhead{$\dm_{84\%}$} & \colhead{$\dm_{95\%}$} & \colhead{$\sigma_{\dm}$} &
\colhead{$M_r$} & \colhead{$\sigma_{M_r}$} & \colhead{$D$} & \colhead{$\sigma_D$} & \colhead{$r_{\rm gc}$} &
\colhead{$\sigma_{r_{\rm gc}}$} \\
\colhead{$M_r$} & \colhead{$\sigma_{M_r}$} & \colhead{$D$} & \colhead{$\sigma_D$} & \colhead{$r_{\rm gc}$} &
\colhead{$\sigma_{r_{\rm gc}}$} \\
\colhead{[deg]} & \colhead{[deg]} & \colhead{[mag]} & \colhead{[mag]} & \colhead{[mag]} &  \colhead{[mag]} & \colhead{[${\rm km\,s^{-1}}$]} & \colhead{[${\rm km\,s^{-1}}$]} & \colhead{[K]} & \colhead{[K]} & \colhead{[dex]} & \colhead{[dex]} & \colhead{[dex]} & \colhead{[dex]} & \colhead{[mag]} & \colhead{[mag]} & \colhead{[mag]} & \colhead{[mag]} & \colhead{[mag]} & \colhead{[mag]} & \colhead{[mag]} & \colhead{[mag]} & \colhead{[kpc]} & \colhead{[kpc]} & \colhead{[kpc]} & \colhead{[kpc]}
} 
\colnumbers
\startdata 
  272.152 &   38.854 & 14.853 &  0.037 & 1.071 & 0.018 & -111.2 &   3.8 & 4137.0 &  41.0 & 1.90 & 0.07 & -0.53 & 0.04 & 15.48 & 15.61 & 15.79 & 15.97 & 16.08 &  0.18 & -0.94 &  0.19 & 14.38 &  1.21 & 13.59 &  1.01 \\ 
 232.282 &   41.214 & 13.324 &  0.037 & 1.161 & 0.017 & -161.6 &   3.9 & 4027.0 &  51.0 & 1.80 & 0.09 & -0.54 & 0.05 & 14.27 & 14.38 & 14.53 & 14.68 & 14.80 &  0.15 & -1.21 &  0.16 &  8.06 &  0.56 &  9.99 &  0.35 \\ 
 187.613 &   29.159 & 14.306 &  0.040 & 0.827 & 0.035 &  -99.9 &   4.8 & 4476.0 &  47.0 & 2.61 & 0.08 & -0.68 & 0.04 & 13.87 & 14.07 & 14.37 & 14.66 & 14.84 &  0.30 & -0.06 &  0.30 &  7.45 &  1.02 & 11.40 &  0.73 \\ 
 214.504 &   14.421 & 13.593 &  0.040 & 1.227 & 0.032 &   45.3 &   3.7 & 3851.0 &  59.0 & 1.77 & 0.10 & -0.38 & 0.06 & 14.44 & 14.57 & 14.74 & 14.88 & 14.97 &  0.15 & -1.13 &  0.16 &  8.82 &  0.63 &  9.22 &  0.38 \\ 
 222.955 &   19.915 & 14.372 &  0.039 & 1.059 & 0.028 & -115.0 &   3.9 & 4119.0 &  42.0 & 1.87 & 0.07 & -0.52 & 0.04 & 14.91 & 15.04 & 15.25 & 15.45 & 15.57 &  0.20 & -0.87 &  0.20 & 11.20 &  1.04 & 10.58 &  0.76 \\ 
\enddata
\end{splitdeluxetable}



\end{document}